\documentclass{article} 
\usepackage{nips12submit_e,times}

\usepackage{amsmath}
\usepackage{amsfonts}
\usepackage{latexsym}
\usepackage[ruled,vlined]{algorithm2e}

\usepackage{graphicx}

\usepackage[all]{xy}

\usepackage[T1]{fontenc} 

\title{Gaussian process regression as a predictive model for Quality-of-Service in Web service systems}
\author{
Jakub M.~Tomczak \\
Wroc\l aw University of Technology \\
Wroc\l aw, Poland \\
\texttt{jakub.tomczak@pwr.wroc.pl} \\
\And
Jerzy \'{S}wi\k{a}tek \\
Wroc\l aw University of Technology \\
Wroc\l aw, Poland \\
\texttt{jerzy.swiatek@pwr.wroc.pl}
\And
Krzysztof Latawiec \\
Opole University of Technology \\
Opole, Poland \\
\texttt{k.latawiec@po.opole.pl}
}

\nipsfinalcopy 

\graphicspath{ {images/} }

\begin{document}

\maketitle

\begin{abstract}
In this paper, we present the \emph{Gaussian process regression} as the predictive model for \textit{Quality-of-Service} (QoS) attributes in Web service systems. The goal is to predict performance of the execution system expressed as QoS attributes given existing execution system, service repository, and inputs, e.g., streams of requests. In order to evaluate the performance of Gaussian process regression the simulation environment was developed. Two quality indexes were used, namely, Mean Absolute Error and Mean Squared Error. The results obtained within the experiment show that the Gaussian process performed the best with linear kernel and statistically significantly better comparing to \emph{Classification and Regression Trees} (CART) method.
\end{abstract}

\section{Introduction}

Performance prediction in web service systems is one of the most important issues in modern computer networks which is still insufficiently solved by well-known methods because of the gap between theoretical considerations and applications. This research explores this issue. In general, web service systems consist of the following layers \cite{TED:09}: (i) an \emph{execution layer} which controls the execution of composite Web services and manages dataflow between them, and (ii) an \emph{application service layer} which delivers requested functionalities to clients. Web services are designed according to service oriented computing (SOC) paradigm \cite{PG:03} and represent encapsulated functionalities of applications. 

In this paper, we focus on the execution layer only. Our goal is to predict performance of the execution system expressed as \textit{Quality-of-Service} (QoS) attributes given existing execution system, service repository, and inputs, e.g., streams of requests. The predicted performance can be used not only for personalization of services \cite{TS:11} but most of all for service selection \cite{PZ:11, SZWZXM:07} and resource allocation \cite{RG:11, WCB:01}. For example, modelling dependency between QoS attributes and streams of requests allows to allocate computational resources in an optimal way. Otherwise other techniques are needed, e.g., change detection methods \cite{RT:11, T:12, TZ:12}. However, a predictive model for QoS attributes can be used as an objective function in an optimization task for resource allocation.

According to above facts the proposition of the predictive model becomes a crucial issue. It can be assumed that the execution system with fixed computational resources and given inputs performs roughly in a deterministic manner. Nevertheless, internal and unknown processes within the execution system introduce random noise and thus the QoS attributes are random variables as well. Hence, a probabilistic model seems to be the best suited in the considered application.

Recently, in the literature of machine learning, a non-parametric regression model called \emph{Gaussian process} was introduced \cite{B:06, M:98, RW:06}. Gaussian processes are considered as one of the most successful regression models applied in many domains, e.g., biosystems \cite{AK:07}, predictive control for chemical plants \cite{LK:07}, hydraulic systems \cite{GL:09}, learning inverted pendulum \cite{DR:09} and non-linear system identification \cite{TDR:09}.

The main contribution of this paper is twofold. First, an application of Gaussian process to predicting performance of the execution system in Web service system is presented. Second, a simulation environment for Web traffic is proposed.

The paper is organized as follows. In Sect. 2 the problem of performance prediction in Web service systems is stated. In Sect. 3 details about Gaussian processes models are outlined. In Sect. 3 the simulation environment is described and experiments are conducted. At the end conclusions are drawn.

\section{Prediction of QoS in Web service systems}

Let $\mathbf{x} \in \mathcal{X}$ denote a $D$-dimensional vector of input variables to the execution system. For example, inputs are total sizes of demands from $D$ classes maintained in queues to the execution system, $\mathcal{X} = \mathbb{R}_{+}^{D}$. Outputs of the execution systems are denoted by $\mathbf{y} \in \mathcal{Y}$ and correspond to QoS attributes, e.g., time spent in the execution system (so called \emph{latency}).\footnote{Multivariate regression model can be treated as a problem of several one-dimensional regression models \cite{B:06} and thus, for further simplicity, we will consider only one output (target variable), i.e., $y$.}

Further, we assume that there exists a dependency between inputs and outputs. However, without knowing processes responsible for generating teletraffic and internal processes governing the execution systems we should consider noise in the model. The dependency between inputs and QoS attributes can be seen as a regression model, hence the target variable $y$ is given by a function $f: \mathcal{X} \rightarrow \mathbb{R}$ and a Gaussian additive noise

\begin{equation}\label{eq:model}
y = f(\mathbf{x}) + \varepsilon,
\end{equation}

where $\varepsilon$ is a zero mean Gaussian random variable with precision $\beta^{-1}$, $\varepsilon \sim \mathcal{N}(\cdot|0,\beta^{-1})$.


The prediction task is to return an output $\hat{y}$ for given new inputs $\hat{\mathbf{x}}$ and $N$ historical observations (data) $\mathcal{D} = \{(\mathbf{x}_{n}, y_{n})\}_{n=1}^{N}$. Because we consider a probabilistic model (\ref{eq:model}), we need to calculate the following predictive distribution

\begin{equation}\label{eq:prediction}
p(\hat{y} | \hat{\mathbf{x}}, \mathcal{D}) = \int p(\hat{y}| \hat{\mathbf{x}}, f, \mathcal{D}) p(f|\mathcal{D}) \mathrm{d}f.
\end{equation}

In order to calculate the predictive distribution we are supposed to give \emph{a priori} distribution of dependencies $f$. As we will see shortly, this is analytically tractable as long as prior is Gaussian.

In the context of Web service systems, we want to predict QoS attributes, e.g., latency, for given inputs. Here we do not consider dynamics of input streams, thus the efficiency of our approach relies on proper formulation of input variables and calculation of target variable.

\section{Gaussian process regression}

\subsection{The model}

The idea of Gaussian processes is to put a prior distribution on function $f$ and learn the dependencies basing on available data \cite{B:06, M:98, RW:06}. However, the Gaussian process is a non-parametric model and thus there is no need to formulate any fixed relationships between inputs and target variable. The non-linear regression model using Gaussian process (called \emph{Gaussian process regression}) is as follows:
\begin{align}
y &= f(\mathbf{x}) + \varepsilon, \nonumber\\
f &\sim \mathcal{GP}(\cdot | 0, k(\mathbf{x}, \mathbf{x}')),\\
\varepsilon &\sim \mathcal{N}(\cdot|0,\beta^{-1}), \nonumber
\end{align}

where $\mathcal{GP}$ denotes the Gaussian process, $k(\cdot, \cdot)$ is the covariance (kernel) function.

Now the predictive distribution (\ref{eq:prediction}) is analytically tractable because prior on $f$ is $\mathcal{GP}$, \emph{likelihood} is Gaussian, and \emph{a posteriori} distribution on $f$ is also a $\mathcal{GP}$.

Let $\boldsymbol{\mathsf{y}}$ denote a column of output observations and $\boldsymbol{\mathsf{f}}$ -- a column of $f(\mathbf{x_{n}})$, $n=1\ldots N$. From the definition of Gaussian process the marginal distribution $p(\boldsymbol{\mathsf{f}})$ is as follows:

\begin{equation}\label{eq:marginalF}
p(\boldsymbol{\mathsf{f}}) = \mathcal{N}(\boldsymbol{\mathsf{f}}|\mathbf{0}, \mathbf{K}),
\end{equation}

where $\mathbf{0}$ -- column of zeros, $\mathbf{K}$ -- Gramm matrix, i.e., $K_{n,m} = k(\mathbf{x}_{n}, \mathbf{x}_{m})$.

Similarly, the distribution of $\boldsymbol{\mathsf{y}}$ conditioned by $\boldsymbol{\mathsf{f}}$ is the following:

\begin{equation}\label{eq:conditioned}
p(\boldsymbol{\mathsf{y}}|\boldsymbol{\mathsf{f}}) = \mathcal{N}(\boldsymbol{\mathsf{y}}|\boldsymbol{\mathsf{f}},\beta^{-1}\mathbf{I}_{N}),
\end{equation}

where $\mathbf{I}_{N}$ is $N\times N$ unit matrix.

The marginal distribution $p(\boldsymbol{\mathsf{y}})$ equals

\begin{equation}\label{eq:conditioned}
p(\boldsymbol{\mathsf{y}}) = \mathcal{N}(\boldsymbol{\mathsf{y}}|\mathbf{0},\mathbf{C}),
\end{equation}

where $\mathbf{C}$ is a matrix such that $C_{n,m} = K_{n,m} + \beta^{-1}\delta_{n,m}$, and $\delta_{n,m}$ is a Kronecker's delta.

Because all distributions are Gaussian, hence the predictive distribution\footnote{Here we use shorthand notation in comparison to equation (\ref{eq:prediction}).} $p(\hat{y}|\hat{\mathbf{x}},\boldsymbol{\mathsf{y}})$ is Gaussian distribution with mean and covariance given by \cite{B:06, RW:06}

\begin{align}
m(\hat{\mathbf{x}}) &= \mathbf{k}^{\top}\mathbf{C}^{-1}\boldsymbol{\mathsf{y}},\label{eq:meanPrediction}\\
\sigma^{2}(\hat{\mathbf{x}}) &= k(\hat{\mathbf{x}}, \hat{\mathbf{x}}) + \beta^{-1} - \mathbf{k}^{\top}\mathbf{C}^{-1}\mathbf{k},\label{eq:variancePrediction}
\end{align}

where $\mathbf{k}$ is a vector with elements $k(x_{n},\hat{x})$, $n=1\ldots N$.

Finally, for given inputs to the execution systems $\hat{\mathbf{x}}$ we have calculated the predictive distribution with mean and variance defined as (\ref{eq:meanPrediction}) and (\ref{eq:variancePrediction}), respectively. Gaussian probability density function has one mode which is in the same time mean value, hence the mean value (\ref{eq:meanPrediction}) is the most probable value for given inputs and the variance (\ref{eq:variancePrediction}) determines its uncertainty.

\subsection{Covariance function}

\begin{figure}[!htp]
\centering
\includegraphics[width=1\textwidth]{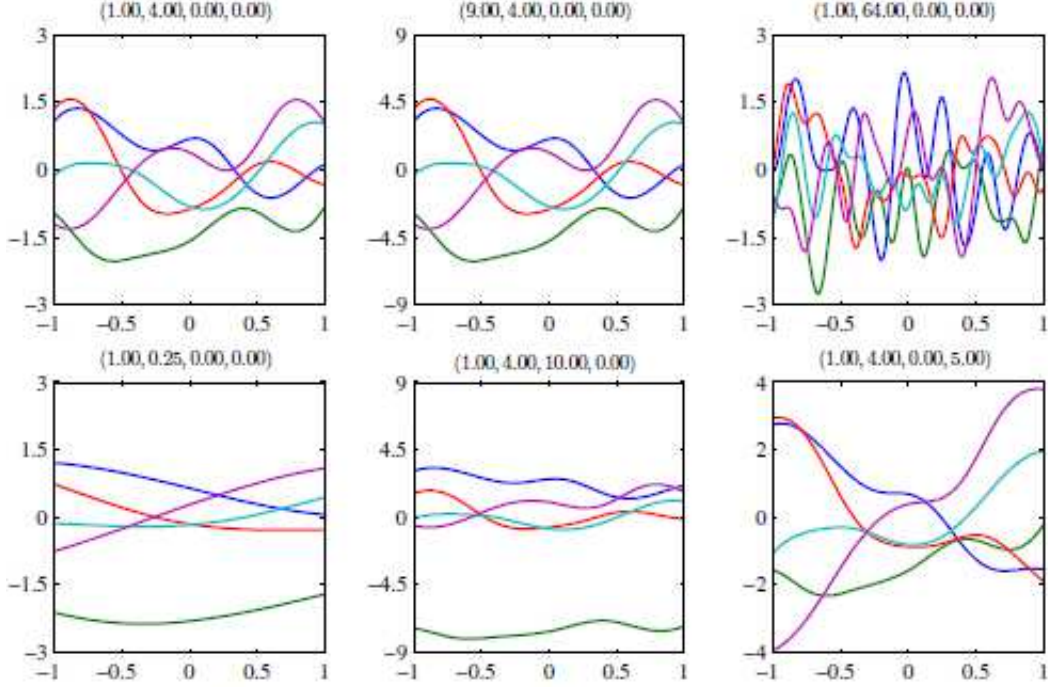}
\caption{Examples from a Gaussian prior defined by the following complex covariance function: $k(\mathbf{x}, \mathbf{x}') = \theta_0 \exp\{-\frac{\theta_1}{2} \|\mathbf{x} - \mathbf{x}'\|^2)\} + \theta_2 + \theta_3 \mathbf{x}^{\top}\mathbf{x}'$. The title above each plot denotes $(\theta_0, \theta_1, \theta_2, \theta_3)$. Figure taken from \cite{B:06}.}
\label{fig:covFun}
\end{figure}

Crucial step in modelling any phenomenon using Gaussian processes is the determination of the kernel function. There are many kernel functions described in literature (see \cite{RW:06} for further details), e.g., linear kernel

\begin{equation}\label{eq:lin}
k_{lin}(\mathbf{x}, \mathbf{x}') = \mathbf{x}^{\top}\mathbf{A}^{-2}\mathbf{x}'
\end{equation}

where $\mathbf{A}^{-2}$ is $D\times D$ diagonal matrix, squared exponential kernel

\begin{equation}\label{eq:se}
k_{se}(\mathbf{x}, \mathbf{x}') = \sigma_{f}^{2} \exp\{(\mathbf{x} - \mathbf{x}')^{\top}\mathbf{A}^{-2}(\mathbf{x} - \mathbf{x}')\}
\end{equation}

where $\sigma_{f}^{2}$ is a bias, and complex kernels, for example,

\begin{equation}\label{eq:seComplex}
k(\mathbf{x}, \mathbf{x}') = k_{se}(\mathbf{x}, \mathbf{x}') + k_{lin}(\mathbf{x}, \mathbf{x}') + b,
\end{equation}

where $b$ is a bias parameter. Choosing specific kernel allows to reflect different similarities between points (see Fig. \ref{fig:covFun}).

\subsection{Learning the hyperparameters}

The predictions of a Gaussian process regression depend mainly on the choice of the covariance function. From the practical point of view it is more convenient to propose a parametric set of covariance functions than to fix the covariance function by hand. Then the inference of the values of hyperparameters can be conducted basing entirely on data. 

In this paper we use the type 2 maximum likelihood procedure which allows to determine hyperparameters values by maximizing the log likelihood function ($\boldsymbol\theta$ denotes a vector of hyperparameters)

\begin{equation}\label{eq:logLikelihood}
\ln p(\boldsymbol{\mathsf{y}}|\boldsymbol\theta) = -\frac{1}{2} \ln |\mathbf{C}| - \frac{1}{2} \boldsymbol{\mathsf{y}}^{\top} \mathbf{C}^{-1} \boldsymbol{\mathsf{y}} - \frac{N}{2}\ln(2\pi).
\end{equation}

If the evaluation of derivatives of $\mathbf{C}$ is straightforward we can easily calculate the following derivatives \cite{B:06}

\begin{equation}\label{eq:logLikelihood}
\frac{\partial}{\partial \theta_{i}}\ln p(\boldsymbol{\mathsf{y}}|\boldsymbol\theta) = -\frac{1}{2} \mathrm{Tr} \Big{(} \mathbf{C}^{-1} \frac{\partial \mathbf{C}}{\partial \theta_{i}} \Big{)} + \frac{1}{2} \boldsymbol{\mathsf{y}}^{\top} \mathbf{C}^{-1} \frac{\partial \mathbf{C}}{\partial \theta_{i}} \mathbf{C}^{-1} \boldsymbol{\mathsf{y}}.
\end{equation}

\section{Experiments}

\subsection{Preliminaries}

The purpose of the experiment is to examine the prediction quality of the proposed approach. We take under consideration a Web service execution environment and average latency of services\rq{} responses in the system as an output QoS attribute. To reflect the nature of real Web service execution system we propose a simulation environment written in Matlab\textsuperscript{\textregistered}. The simulation model is presented in Fig. \ref{fig:simEnvironment}. The model consists of the following components: (i) \emph{teletraffic generator} (TG), which imitates clients\rq{} behaviour by generating Web service requests, (ii) \emph{scheduler}, which distributes service requests to proper queues, (iii) \emph{queues}, which maintain demands and work in FIFO fashion,(iv) \emph{Round Robin} (RR), which collects requests from queues in circular order \cite{SGG:98}, (v) \emph{execution system}, which executes the services.

\begin{center}
\begin{figure}[!htp]
$$\xymatrix{ & & *+[F-:<3pt>]{\textrm{Queue 1}} \ar@{->}[ddr] & & & \\
			 & & *+[F-:<3pt>]{\textrm{Queue 2}} \ar@{->}[dr] & & & \\
			*++[F-,]{\textrm{TG}} \ar@{->}[r] & *++[o][F-]{\textrm{Scheduler}} \ar@{->}[uur] \ar@{->}[ur] \ar@{->}[r] \ar@{->}[dr] & \ldots \ar@{->}[r] & *++[o][F-]{RR} \ar@{->}[r] & *++[F-,]{\textrm{ES}} \ar@{->}[r] & *+[F-]{\textrm{Sink}} \\
			 & & *+[F-:<3pt>]{\textrm{Queue D}} \ar@{->}[ur] & & & }$$
\caption{Schematic diagram of the simulation environment, TG -- teletraffic generator, RR -- Round Robin, ES -- execution system.}
\label{fig:simEnvironment}
\end{figure}
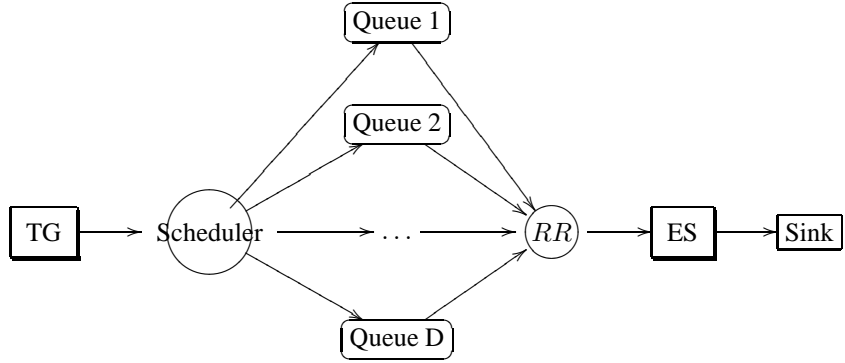
\end{center}

The simulation environment provides teletraffic: inputs $\mathbf{x}$ to the execution system which are sizes of queues, and outputs of the execution systems which are average latencies $y$. Each quantity is calculated using $T$ last observations. In the experiment we compare Gaussian process regression with well-known \emph{Classification And Regression Trees} (CART) method \cite{BFOSSC:83} which is a baseline in the experiment. We use Matlab implementation of CART and a toolbox for Gaussian processes provided by Rasmussen and Nickisch \cite{RN:10}.

\subsection{Simulation details}

\subsubsection{Modelling teletraffic}

We assume that each demand to the system can occur with a probability $p$. Then the class of the demand is generated with uniform probability and the size of the demand is drawn from lognormal distribution.\footnote{According to \cite{BC:98}, probability distribution function for the size of the file body is lognormal.} The process of generating a demand is as follows (assuming some universal time unit, e.g., one second):
\begin{enumerate}
\item Generate random number from interval $[0,1]$. If it is greater than $p$, then go to step 2. Otherwise go to step 1.
\item Generate the demand class using uniform probability.
\item Generate the size of the demand using lognormal distribution. Go to step 1.
\end{enumerate} 

\subsubsection{Modelling execution system}

One demand arrives to the execution system according to the Round Robin scheduler. Execution of a demand takes as many universal time units as it is completely executed, i.e., the executed size of the demand is zero. In the simulation environment we need to determine \emph{execution sizes} per one universal time unit for each demand class.

\subsubsection{Prediction and evaluation}

In order to evaluate Gaussian process regression and CART we have to determine a number of training points (inputs with outputs) and a number of test points. We use the following quality indexes:
\begin{itemize}
\item Mean Absolute Error (MAE);
\item Mean Squared Error (MSE).
\end{itemize}
We allow three covariance functions in the simulation environment, i.e., linear kernel (\ref{eq:lin}), squared exponential kernel (\ref{eq:se}) and complex kernel expressed as in equation (\ref{eq:seComplex}). Moreover, we use Gaussian likelihood with precision $\beta^{-1}$ (called \emph{noise} in the simulation environment), and type 2 maximum likelihood procedure for hyperparameters learning.

\begin{figure}[!htp]
\centering
\includegraphics[width=1\textwidth]{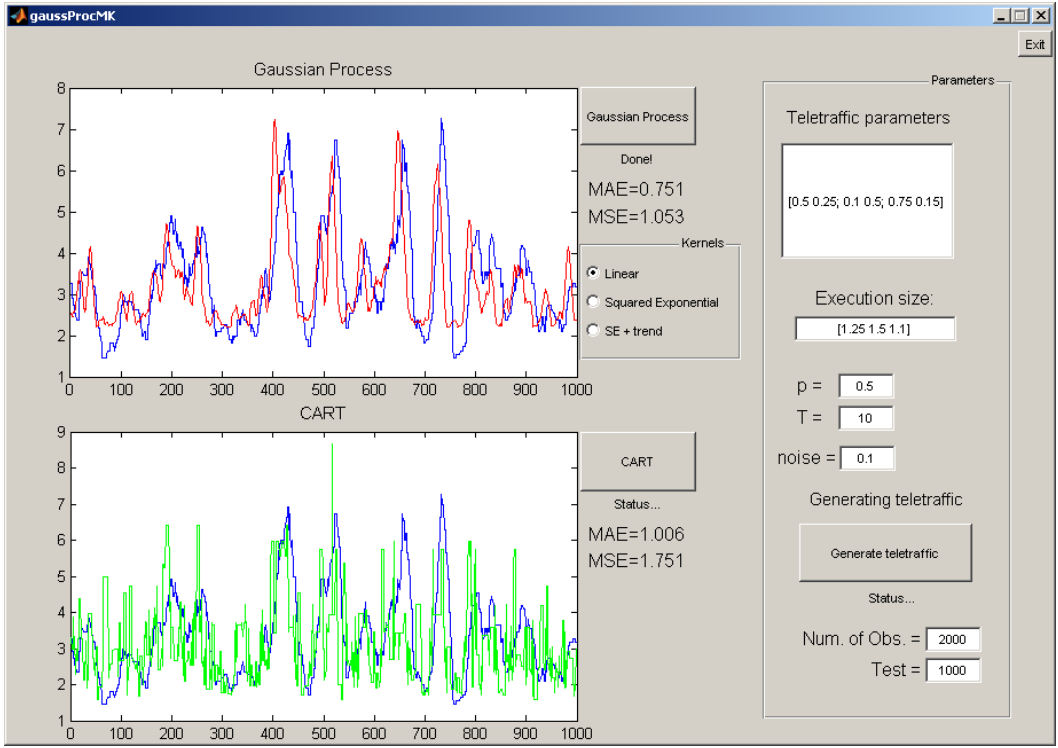}
\caption{GUI of the simulation environment.}
\label{fig:gui}
\end{figure}

\subsection{Results and discussion}

The GUI of the simulation environment (see Fig. \ref{fig:gui}) allows to fix all parameters. In order to conduct experiments we generated $10$ simulations of teletraffic with the following parameters:
\begin{itemize}
\item number of classes -- $3$;
\item number of test points -- $1000$;
\item number of training points -- $1000$;
\item $p = 0.5$;
\item $T = 10$;
\item $\beta^{-1} = 0.1$;
\item parameters of lognormal distribution -- i) for class 1: $0.5, 0.25$, ii) for class 2: $0.1, 0.5$, iii) for class 3: $0.75, 0.15$;
\item execution size per one universal time unit: i) for class 1: $1.25$, ii) for class 2: $1.5$, iii) for class 3: $1.1$.
\end{itemize}

\begin{figure}[!htp]
\centering
\includegraphics[width=1\textwidth]{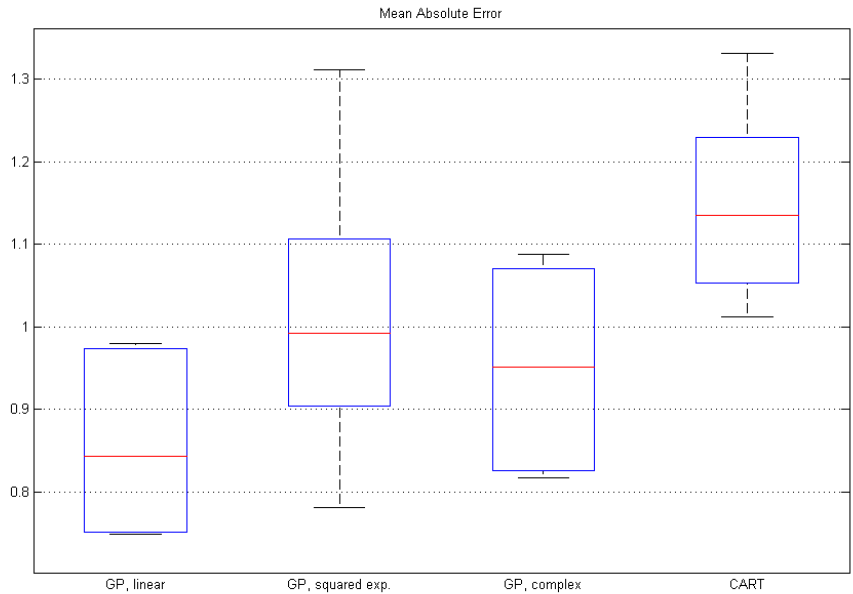}
\caption{Box-and-whisker plot for MAE results. Red line represents median value, blue box -- quantiles, and black lines -- range of values.}
\label{fig:boxplotMAE}
\end{figure}

\begin{figure}[!htp]
\centering
\includegraphics[width=1\textwidth]{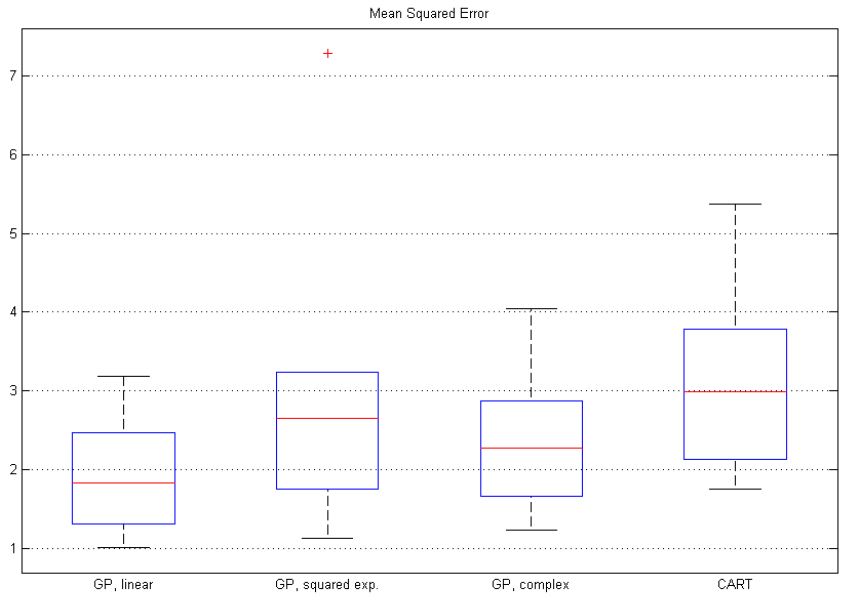}
\caption{Box-and-whisker plot for MSE results. Red line represents median value, blue box -- quantiles, black lines -- range of values, and red cross -- outlier.}
\label{fig:boxplotMSE}
\end{figure}

The results of $10$ simulations are gathered and represented as a box-and-whisker plot in Fig \ref{fig:boxplotMAE} and \ref{fig:boxplotMSE} for MAE and MSE, respectively. The Gaussian process regression performed better in comparison to CART for any kernel function. However, the linear kernel appeared to be more proper to represent similarity between inputs than squared exponential or complex kernel. Especially, the complex kernel is a sum of linear and squared exponential kernels and that is why it performed slightly better than squared exponential.

In order to compare the results, we also performed two sample $t$-test for MAE at the $5\%$ significance level between Gaussian process regression with linear kernel and CART. The null hypothesis, i.e., random samples share the same mean and equal but unknown variances, can be rejected with $p$-value equal $1.8406\times 10^{-5}$. Similarly, for MSE, we can reject the null hypothesis with $p$-value equal $0.018$. In other words, the Gaussian process regression performs statistically better than CART.

\newpage
\section{Conclusions}

In this paper, we have presented the Gaussian process regression as the predictive model for QoS attributes in Web service systems. The idea of Gaussian process regression is well-grounded in the field of machine learning but its application in Web service systems is novel. In order to evaluate the performance of Gaussian process regression the simulation environment was developed. Two quality indexes were used, namely, Mean Absolute Error and Mean Squared Error. The results show that the Gaussian process performed the best with linear kernel and statistically better comparing to CART method.

The proposed approach shows that application of machine learning methods can develop existing computer network systems. The results presented in the paper indicate high accuracy but further research and experiments, especially on existing systems, are necessary. Summing up, this paper tries to fill the gap between machine learning methods and computer network applications.

\subsubsection*{Acknowledgments}
The research conducted by Jakub M. Tomczak has been partially co-financed by the European Union within the European Social Fund.

\bibliographystyle{abbrv}
\bibliography{nips}

\end{document}